\documentclass{PoS}
\usepackage{amssymb}
\usepackage{gensymb}
\usepackage{graphicx}
\usepackage{amsmath}
\usepackage{wasysym}
\usepackage{verbatim}

\title{Constraints on the Diffuse Gamma-Ray Background with HAWC}

\ShortTitle{HAWC DGB}

\author{\speaker{J. Patrick Harding}$^{\rm a}$
  {for the HAWC Collaboration}\footnote{For a complete author list, see http://www.hawc-observatory.org/collaboration/icrc2019.php}\\
        $^{\rm a}$Los Alamos National Laboratory\\
        E-mail: \email{jpharding@lanl.gov}}

\abstract{The Diffuse Gamma-Ray Background (DGB) above 100 GeV at high-latitudes is expected to be produced by unresolved extragalactic objects such as active galactic nuclei, isotropic Galactic gamma-rays, and possible emission from dark matter annihilations or decays in the Galactic dark matter halo. The DGB has been observed up to nearly 1 TeV, but has yet to be detected at higher energies. Detections or stringent limits on the DGB above this energy would have strong multimessenger consequences, such as constraining the origin of astrophysical neutrinos observed in IceCube. The High Altitude Water Cherenkov (HAWC) observatory has its highest sensitivity to gamma rays above 1 TeV and observes over 2/3 of the sky each day. With its high energy reach and large areal coverage, HAWC can significantly improve searches of the DGB at TeV energies. We will investigate parameter cuts to the HAWC dataset to better isolate gamma-ray air showers from background hadronic showers. This, combined with new background estimation techniques will improve the HAWC sensitivity to the DGB. We will present a limit on the DGB with HAWC as well as its implications for multimessenger and dark matter studies.}

\FullConference{36th International Cosmic Ray Conference -ICRC2019-\\
		July 24th - August 1st, 2019\\
		Madison, WI, U.S.A.}

\begin{document}

\section{Introduction}
Since the first measurement of the diffuse gamma-ray background (DGB) by the SAS 2 experiment in 1975~\cite{1975ApJ...198..163F}, the source of this emission has remained an open question in astrophysics. With the measurements of subsequent experiments, the DGB has been shown to extend from photon energies of a few MeV up to hundreds of GeV~\cite{Ackermann:2014usa}. The DGB is characterized as being diffuse, isotropic emission which is uncorrelated with any known emission source. It follows a nearly featureless power-law in energy over five orders of magnitude, possibly cutting off above 1 TeV. Understanding the source of the DGB is our best way to understand the nature of ultrafaint astrophysical objects.

Most hypothesized sources of the DGB are extragalactic in nature, such as unresolved active galactic nuclei and starburst galaxies. However, galactic contributions to the DGB may also exist -- particularly if the DGB is observed to multi-TeV energies where extragalactic emission is attenuated via pair-production on the extragalactic background light. Therefore, extragalactic sources are unlikely to dominate the DGB at high energy. A Galactic source of gamma-rays, such as galactic dark matter annihilations, is a much more likely candidate for any observed high-energy DGB emission.

The High Altitude Water Cherenkov (HAWC) gamma-ray observatory is a wide field-of-view observatory sensitive to gamma-rays and cosmic-rays at energies from 300 GeV to \textgreater100 TeV. The HAWC design consists of 300 water Cherenkov detectors (WCDs) each observing particles in air showers initiated by high-energy astrophysical particles. Here, we use the large sky coverage of HAWC to search for evidence of the DGB above 10 TeV. In particular, we will be using 815 days of HAWC data, taken from November 2014 -- February 2017. In the highest HAWC energy bin used, where $\ge84\%$ of the WCDs are hit, the initiating gamma rays come from energies of 20-160 TeV. This is the energy range of the data used in this analysis.

\section{HAWC Analysis at High Signal-to-Noise}\label{anal}
\begin{figure*}[t!]
\makebox[\textwidth][c]{
\begin{tabular}{cc}
\includegraphics[width=.5\textwidth]{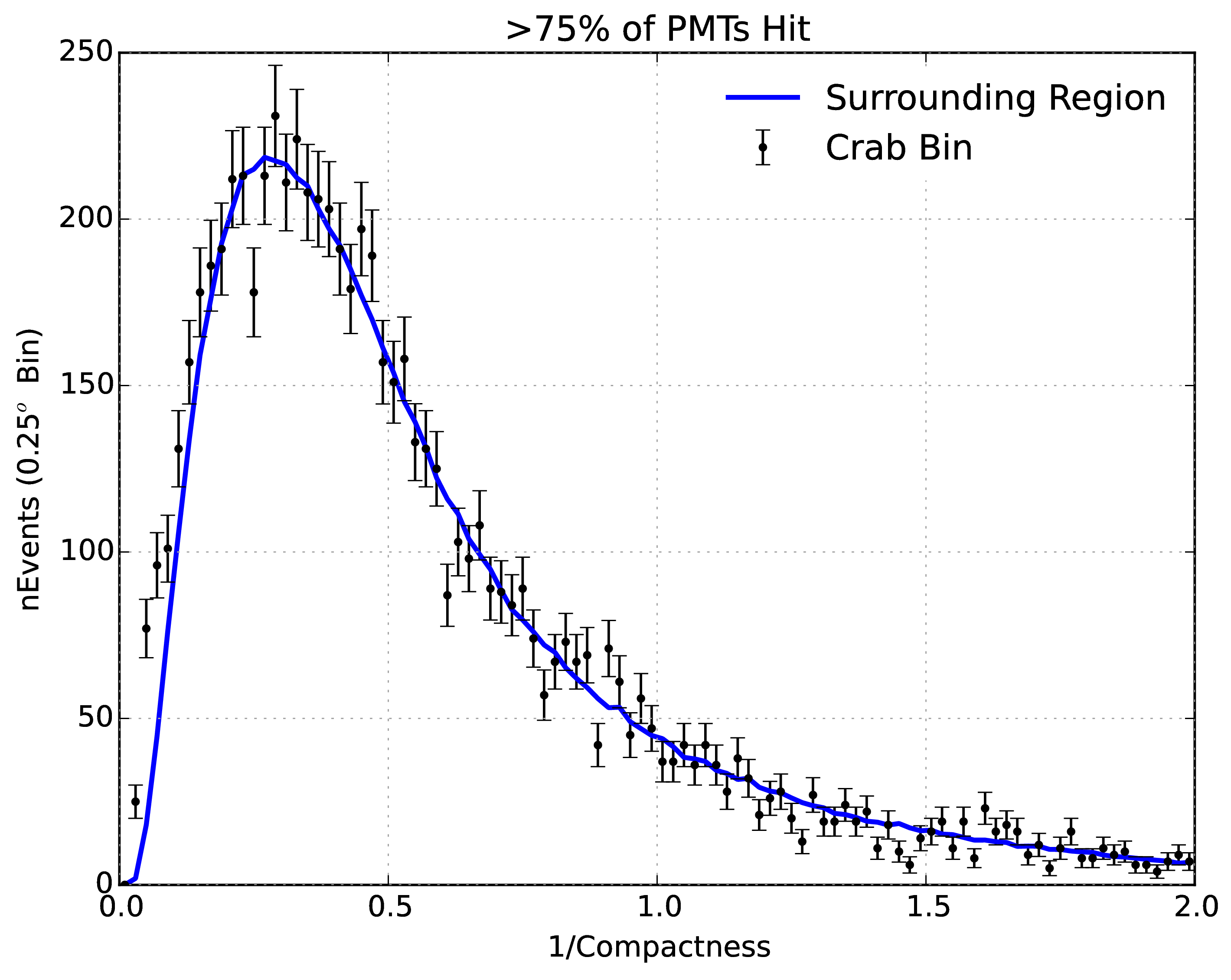}
\includegraphics[width=.5\textwidth]{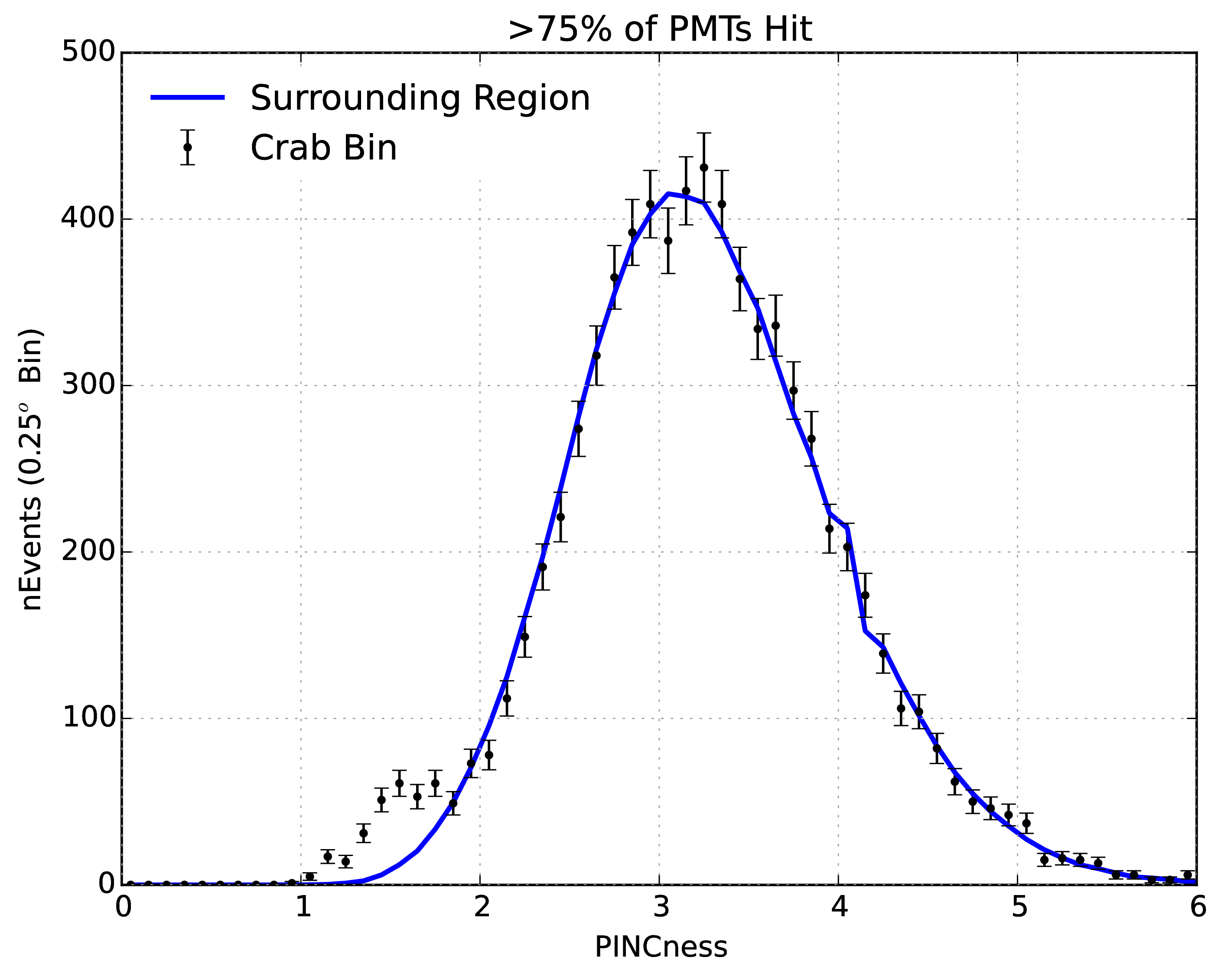}
\end{tabular}
}
\caption[caption]{\label{compPINC} Distribution of events around the Crab nebula for different values of $1/$Compactness (left) and PINCness (right). The blue line shows the distribution in a region around the Crab nebula which is expected to only contain hadronic cosmic rays, while the black data points indicate the distribution in a region which includes the Crab nebula itself and is therefore expected to have a large gamma-ray population in addition to cosmic-ray backgrounds. In both distributions, events with lower values are significantly more likely to be gamma rays than hadronic cosmic rays. }
\end{figure*}

As with all HAWC analyses, the search for DGB signals in HAWC relies on removing large astrophysical cosmic ray backgrounds from the data while maintaining as much of the desired gamma-ray signal as possible. Even the brightest source in the HAWC sky, the Crab nebula, only emits around one-hundred TeV-scale gamma rays per day, while the total HAWC background rate is $\sim 25$ kHz~\cite{Abeysekara:2017mjj}. To optimize its signal-to-background, HAWC currently uses two variables to separate gamma-ray showers from hadronic showers -- PINCness and Compactness. Both of these parameters account for the relative smoothness of an electromagnetic air shower across the HAWC array compared to a hadronic shower. Shown in figure~\ref{compPINC}, cosmic ray backgrounds and Crab nebula data look similar at large values of PINCness and $1/$Compactness. However, for low values of these variables, the data from the Crab nebula shows an excess over cosmic ray backgrounds. Cutting out data with large values of PINCness and $1/$Compactness efficiently removes cosmic ray backgrounds from the data while retaining gamma rays. For more information on these parameters, see Ref.~\cite{Abeysekara:2017mjj}. It should also be noted that due to the similarity of air showers induced by gamma rays and electrons (positrons), any electron (positron) air showers are as likely to pass these cuts as the gamma-ray showers. 

The difficulty with searching for the DGB is that it exists in all directions and therefore has no morphology to leverage in distinguishing from cosmic ray backgrounds. 
Because the DGB covers the whole sky, any isotropic signal (including cosmic hadrons) has a similar spatial distribution. Because of this, the DGB analysis uses much harder cuts on the PINCness and Compactness variables to remove as much background as possible. As the cuts get harder, the number of total gamma rays expected reduces but the signal-to-background ratio improves. 

\begin{figure*}[t!]
\makebox[\textwidth][c]{
\begin{tabular}{cc}
\includegraphics[width=.75\textwidth]{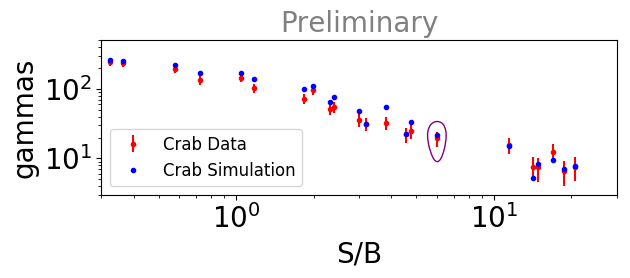}
\end{tabular}
}
\caption[caption]{\label{StoB} The number of gamma rays observed within $1.5\degree$ of the Crab nebula in data (red) and in simulation (blue) for increasing signal-to-background cuts. The signal-to-background ratio here is calculated as the number of simulated gamma rays within $1.5\degree$ of the Crab nebula to the observed background in a similarly-sized region with no known gamma-ray sources. Even over a large range of increasingly hard cuts, the data and simulation agree within uncertainties. The purple circle indicates the selected signal-to-background chosen in this analysis.}
\end{figure*}

To test the performance of the HAWC simulation at these high signal-to-background ratios, we compare the expected gamma-ray signal on the Crab nebula for data and simulation. In this work, we run simulation using the best-fit HAWC Crab spectrum~\cite{Abeysekara:2019edl}
\begin{equation}
\frac{dN}{dE}=2.35\times10^{-13}\left(\frac{E}{7\, \rm TeV}\right)^{-2.79-0.10\ln(E/7\rm\, TeV)}\,\rm (TeV\,cm^2\,s)^{-1}\enspace.
\end{equation}
For increasingly hard cuts on Compactness and PINCness, we calculate the number of gamma rays observed within $1.5\degree$ of the Crab nebula in data and in simulation. This is plotted in figure~\ref{StoB}. 

As expected, there are fewer gamma rays surviving the quality cuts at high signal-to-background. However, the agreement between the observations and simulations are within error bars over the full range of cuts considered. For this analysis, we will choose our cuts as PINCness$\le1.25$ and $1/$Compactness$\le0.079$. This choice of cut gives a high signal-to-background ratio of 6.0 while retaining nearly 20 events for the analysis.

\section{HAWC Limits on the Diffuse Gamma-ray Background}\label{limits}

 For our analysis, we chose a fairly large ``On'' region of $1.5\degree$ around the Crab nebula for comparison with simulation. This choice of region is much larger than the HAWC point-spread function above 20 TeV and therefore should contain nearly all of the photons from the Crab. Additionally, such a large region minimizes any possible discrepancies between the HAWC point-spread function in data and simulation. For the ``Off'' region, we consider an annulus from $2\degree$-$4\degree$ around the Crab nebula. This region is far enough from the Crab nebula to minimize any gamma-ray contamination but close enough to have similar detector-response. That is, the verification of gamma-rays on the Crab nebula should be a good verification of the HAWC effective area to gamma rays at the strong cuts of this analysis.

\begin{figure*}[t!]
\makebox[\textwidth][c]{
\begin{tabular}{cc}
\includegraphics[width=.75\textwidth]{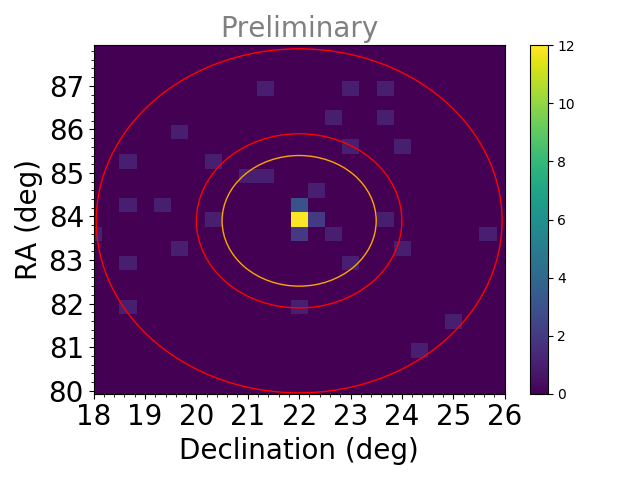}
\end{tabular}
}
\caption[caption]{\label{map} Map of the location of events within $4\degree$ of the Crab nebula surviving the analysis cuts. The orange circle indicates the $1.5\degree$ radius used to define the Crab nebula region. The data between the red circles at $2\degree$ and $4\degree$ is the 19 events which should only be due to DGB photons and residual isotropic backgrounds.}
\end{figure*}

 With the choice of cuts described above and in section~\ref{anal}, we find
\begin{itemize}
\item 19.4 gamma rays within $1.5\degree$ of the Crab nebula (23 counts with 3.6 background)
\item 19 events in a region from $2\degree$-$4\degree$ around the Crab (0.011sr), which should only contain DGB photons and isotropic backgrounds
\item 21.5 expected gamma rays from the Crab Nebula from simulation
\end{itemize}
A spatial map of these events is shown in figure~\ref{map}.

Given the observation in this 0.011sr region, we can constrain the maximum possible flux from the DGB at these energies. For 19 observed counts with Poisson uncertainties, the 95\% one-sided upper limit on the possible counts from the DGB is 26.2 ($\Delta\chi^2=2.71$). Assuming a Crab-like spectrum, this limits the maximum flux from the DGB integrated over solid angle to be $1.22\times$ the Crab flux, or
\begin{equation}
\left(\frac{dN}{dE}\right)_{95\%\rm\, CL}=2.86\times10^{-13}\left(\frac{E}{7\, \rm TeV}\right)^{-2.79-0.10\ln(E/7\rm\, TeV)}\,\rm (TeV\,cm^2\,s)^{-1}\enspace.
\end{equation}
Accounting for the solid angle of the observation, this yields a limit of
\begin{equation}\label{limit}
\left(\frac{d^2N}{dEd\Omega}\right)_{95\%\rm\, CL}=2.60\times10^{-11}\left(\frac{E}{7\, \rm TeV}\right)^{-2.79-0.10\ln(E/7\rm\, TeV)}\,\rm (TeV\,cm^2\,s\,sr)^{-1}\enspace.
\end{equation}
This is the strongest limit on the DGB in the 20-160 TeV energy range.

\begin{figure*}[t!]
\makebox[\textwidth][c]{
\begin{tabular}{cc}
\includegraphics[width=.75\textwidth]{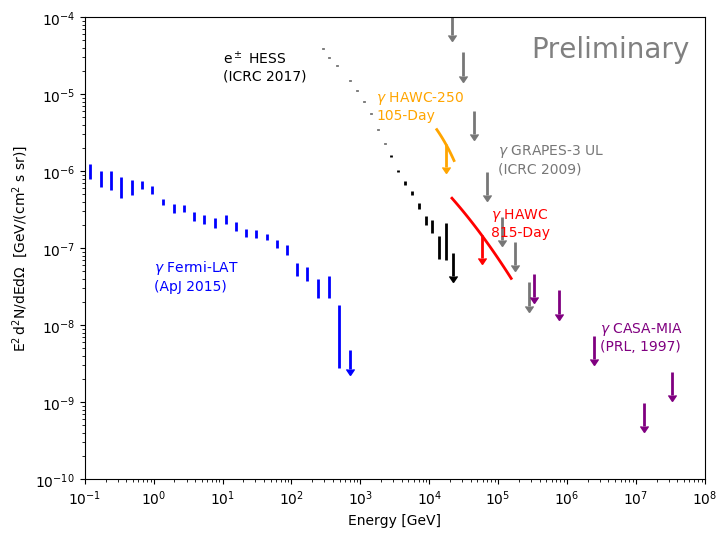}
\end{tabular}
}
\caption[caption]{\label{limitelectrons} The limit on the DGB from 815 days of HAWC data (red) compared to the diffuse electron/positron flux observed by HESS (black)~\cite{HESSICRC1,HESSICRC2}
The observed DGB by the Fermi-LAT (blue)~\cite{Ackermann:2014usa} is shown, as well as previous high-energy limits by GRAPES (gray)~\cite{GRAPES} and CASA-MIA (purple)~\cite{Chantell:1997gs}. The ``HAWC-250 105-Day'' limit (orange) comes from an earlier configuration of HAWC, with significantly less data~\cite{Pretz:2015wma}. The HAWC limit comes within a factor of 3 of the electron/positron spectrum.}
\end{figure*}

\begin{figure*}[t!]
\makebox[\textwidth][c]{
\begin{tabular}{cc}
\includegraphics[width=.75\textwidth]{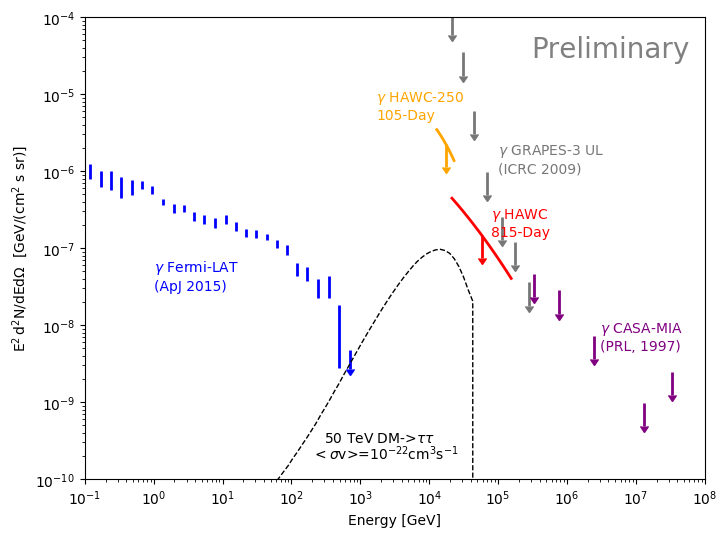}
\end{tabular}
}
\caption[caption]{\label{limitDM} The limit on the DGB from 815 days of HAWC data (red) compared to the diffuse gamma-ray emission from galactic dark matter annihilation into tau leptons (black dashed). The observed DGB by the Fermi-LAT (blue)~\cite{Ackermann:2014usa} is shown, as well as previous high-energy limits by GRAPES (gray)~\cite{GRAPES} and CASA-MIA (purple)~\cite{Chantell:1997gs}. The ``HAWC-250 105-Day'' limit (orange) comes from an earlier configuration of HAWC, with significantly less data~\cite{Pretz:2015wma}. The HAWC limit comes within a factor of 3 of the dark matter spectrum.}
\end{figure*}

\section{Discussion and Prospects}
The flux calculated in equation~\ref{limit} provides a 95\% upper limit on the isotropic fluxes observed by HAWC. While this does include the DGB, it also may include other isotropic signals. To be conservative, we assume that all the flux could be due to DGB, but there are other possible sources of such emission. In particular, one possible component of this flux could be hadrons which appear gamma-like in the HAWC detectors. This can happen, e.g., if a proton interacts in the atmosphere to produce only neutrinos, electrons, positrons, and gamma rays. Such an event could also occur if any muons or other high-transverse-momentum parts of the shower land outside the HAWC array. 

Another possible source of isotropic emission could be cosmic-ray electrons and positrons. At energies above a few TeV, space-based experiments like AMS~\cite{Aguilar:2019ksn} and CALET~\cite{Adriani:2018ktz} have difficulties measuring the low flux of cosmic ray electrons. With much larger effective area, ground-based observatories are better suited to searching for these signals in the TeV range. The HESS observatory has detected an isotropic cosmic-ray flux of electrons and positrons up to 20 TeV~\cite{HESSICRC1,HESSICRC2}. The high HAWC sensitivity at high energies may be able to extend these observations another order of magnitude in energy. The upper limits discussed in section~\ref{limits} already are pushing into an energy range higher than the HESS observation. A comparison of the HESS electron observations to the limits here is shown in figure~\ref{limitelectrons}.

A more exotic possibility is that diffuse emission could be due to annihilating dark matter in the galaxy. Because the Earth is situated in the galaxy, it is sitting near the middle of a cloud of galactic dark matter. If the dark matter annihilates to produce gamma rays, then these gamma rays would be observed in all directions, as a background to all other gamma-ray observations. Therefore, diffuse gamma-ray emission like the DGB at high energies could be indicative of these dark matter annihilations. An example of the kind of diffuse gamma-ray signal expected from dark matter, calculated based on the equations in Ref.~\cite{Abazajian:2010zb}, are seen in figure~\ref{limitDM}.

Overall, the future prospects of the HAWC search for the DGB at TeV energies are bright. This study only considered a relatively small angular region on the sky, 0.011sr. However, the full HAWC sky is larger than 8sr. With this large angular region, even more stringent signal-to-background cuts should still give a reasonable sample size, but with much cleaner data. This analysis is underway. With this improved signal-to-background, HAWC should be able to observe isotropic signals or place strong astrophysical constraints. Whether observing the electron spectrum, dark matter annihilations, or the DGB, HAWC data at high signal-to-noise should provide a unique view into the gamma-ray sky.

\bibliographystyle{ICRC}
\bibliography{references}



\end{document}